\def\a{\alpha}

\def\ve{\varepsilon}

\def\a{\alpha}

\def\IR{\relax{\rm I\kern-.18em R}}
\font\cmss=cmss10 \font\cmsss=cmss10 at 7pt
\def\IZ{\relax\ifmmode\mathchoice
{\hbox{\cmss Z\kern-.4em Z}}{\hbox{\cmss Z\kern-.4em Z}}
{\lower.9pt\hbox{\cmsss Z\kern-.4em Z}}
{\lower1.2pt\hbox{\cmsss Z\kern-.4em Z}}\else{\cmss Z\kern-.4em
Z}\fi}
\def\IN{\relax{\rm I\kern-.18em N}}

\documentstyle[aps]{revtex}

\newcommand{\bge}{\begin{equation}}
\newcommand{\ege}{\end{equation}}
\newcommand{\bga}{\begin{eqnarray}} 
\newcommand{\ega}{\end{eqnarray}}
\newcommand{\nnu}{\nonumber}
\begin{document}
\draft
\title{Coexistence of Band Jahn Teller Distortion and
superconductivity in correlated systems}
\author{Haranath Ghosh}
\address{Instituto de Fisica, Universidade Federal Fluminense,
Campus
da praia Vermelha, Av. Litoranea s/n, 24210 - 340, Niteroi RJ,
Brazil.}
\author{Manidipa Mitra and S. N. Behera}
\address{Institute of Physics, Sachivalaya Marg, Bhubaneswar -751
005, INDIA.} 
\author{S. K. Ghatak}
\address{Department of Physics and Meteorology, Indian Institute of
Technology,
Kharagpur - 721 302, INDIA.}
\maketitle
\begin{abstract}
The co-existence of band Jahn-Teller (BJT) effect with 
superconductivity (SC) is studied for correlated systems, with
orbitally degenerate bands
using a simple model. The Hubbard model for a doubly degenerate
orbital with the on-site intraorbital Coulomb repulsion treated in 
the slave boson formalism and the interorbital Coulomb repulsion
treated in the Hartree-Fock mean field approximation, describes the
correlated system. The model further incorporates the BJT
interaction and a pairing term to account for the lattice
distortion and superconductivity respectively. It is found that
structural 
distortion tends to suppress superconductivity and when SC sets in
at low temperatures, the growth of the lattice distortion
is arrested. The phase diagram comprising of the SC and structural
transition temperatures $T_c$ and $T_s$ versus the dopant
concentration 
$\delta $ reveals that the highest obtainable $T_c$ for an optimum
doping is limited by structural transition. The dependence of the 
occupation probabilities of the different bands as well as the density of states (DOS) in the
distorted-superconducting  
phase, on electron correlation has been discussed. 
\end{abstract}
\maketitle
{\bf PACS. No. : 74.25.DW ; 74.20.Fg ; 76.62.-c } 

The dependence of $T_c$ on structural properties is rather less understood
and had been a crucial question since the discovery of the high
temperature
superconductivity (HTSC) in the cuprates. On the other hand, the dominant effect of electron 
correlation on the normal and SC properties of the cuprates had been realized soon after the
discovery \cite{1}. The structural transitions associated with the 
crystallographic symmetry has been observed in many intermetallic 
compounds \cite{2} as well as in the high $T_c$ cuprates \cite{3}
and 
fullerides \cite{4}. A common feature to all these systems is the
existence
of the Fermi level (FL) in an orbitally degenerate band.
The essence of the Band Jahn Teller (BJT) distortion is the following.
If the Fermi level of a system lies in an orbitally degenerate band then there
can be a net gain in electronic energy if the degeneracy is removed by the 
lowering of the lattice symmetry due to production spontaneous strain which 
results in a structural transition. The production of strain will cost elastic
energy, therefore, the strain will be stabilized if the gain in the electronic energy overcomes the cost in elastic energy. The system will like to optimize 
the distortion so as to maximize the gain in energy, which will happen when the
lower band is maximally occupied and the upper split band is minimally 
occupied. The system would however remain metallic in case of a small but 
finite distortion, as the two splited bands can overlap. 
On the other hand, a split
band with a gap and a resultant insulating phase would arise because of large
distortion. On deviating from the half-filled situation by doping holes 
into the system will result in lowering of the band filling, thereby reducing 
the gain in energy, hence  the magnitude of strain is expected to reduce. Thus
the difference in the occupation probability of the two bands will couple
to the lattice strain which defines the Jahn-Teller Hamiltonian. However, 
the redistribution of the density of state associated with Jahn-Teller 
distortion is expected to affect the occurance of superconductivity in 
the system. A drastic reduction in the density of states at the FL due to 
BJT distortion would suppress superconducting transiotion. On the other hand,
if the superconducting pairing is phonon mediated one can expect an enhancement
in the transition temperature because the distortion of the lattice will
soften the frequency of some phonons resulting in an enhancement of the 
coupling conatant.  In this paper therefore, we present a model study of the 
coexistence of BJT distortion and superconductivity in narrow band systems.

The electron correlation is known to play a significant role in
shaping the physical properties of the normal and SC states
of the correlated systems. It is believed that the single band
Hubbard model contains all the necessary physical
ingredients to describe electron correlation in narrow band systems. One of the
recently contrived methods to solve the Hubbard model for any
filling and any value of correlation ($U$) is the slave -
boson technique, as formulated by Kotliar and
Ruckenstein (KR) \cite{5} which introduces four auxiliary boson
fields corresponding to the occupancy of a site, namely empty
($e_i$), doubly occupied ($d_i$), singly occupied ($p_{i\sigma}$/
$p_{i-\sigma}$) with spin $\pm \sigma$ respectively. In terms
of the slave boson field operators the single band Hubbard model
takes the form
\bga
H & = &\sum _{ij, \sigma} t_{ij} z_{i
\sigma}^{\dagger}c_{i\sigma}^{\dagger} c_{j\sigma}z_{j\sigma} + 
U\sum_{i}d_{i}^{\dagger}d_i -
\mu\sum_{i\sigma}c_{i\sigma}^{\dagger}c_{i\sigma} +
\sum_{i}\lambda_{i}(1-e_i^{\dagger}e_{i} 
\nnu \\
& & 
- d_{i}^{\dagger}d_{i}-\sum_{\sigma}
p_{i\sigma}^{\dagger}p_{i\sigma})+
\sum_{i\sigma}\lambda^{\prime}_{i
\sigma}(c_{i\sigma}^{\dagger}c_{i\sigma}
-d_{i}^{\dagger}d_i-p_{i\sigma}^{\dagger}p_{i\sigma})
\ega
where 
$z_{i\sigma}=(1-d_i^{\dagger}d_i-p_{i\sigma}^{\dagger}p_{i\sigma}
)^{-1/2}
(e_i^{\dagger}p_{i\sigma}+p_{i-\sigma}^{\dagger}d_i)(1-e_{i}^{\dagger}e_i
-p_{i-\sigma}^{\dagger}p_{i-\sigma})^{-1/2}$
and the boson field operators are not independent of each other
but are constrained by the requirements of completeness and
local charge conservation at a site, hence the fourth and fifth
terms are added in the Hamiltonian with the help of Lagrange
multipliers ($\lambda_i$ and $\lambda^{\prime}_i$).
The values of the boson field operators and Lagrange multipliers
are determined by minimizing the free energy in the saddle point
approximation. This approximation diagonalizes the $U$ term
whereas renormalizes the hopping term as $\tilde q t_{ij}$ with
$\tilde q = z^{\dagger}z$ ; $\tilde q$ usually being a complicated
function of the reduced correlation $u$ ($={U\over U_c}$), and the dopant concentration
($\delta
$). In this approach solutions are obtained for the
paramagnetic state for all values of $u$ and band fillings
that reproduces the correct Brinkmann-Rice result for
metal-insulator transition at a critical value of correlation ($U_c$) at half-filling. Approximately,
in
the weak correlation limit $\tilde q = 1 - u^2$ and in case of
strong correlation and small values of $\delta$, $\tilde q
={2\delta\over \sqrt{1-u^{-1}}}$ (for details please see ref.
\cite{5,6}).

We generalize the above formalism to a model system in which the electrons are in a
two-fold degenerate $e_g$ band and interact with the lattice as
well as between themselves, the later via a BCS kind of pairing
interaction mediated by some boson exchange. The Hamiltonian for
such a system can be written as, 
\bga
H & = &  \sum_{k \sigma \alpha = 1,2} [ \{\tilde q \epsilon_k -
\mu +2 U^\prime <n_\beta> \} n_{k \sigma \alpha} -(-1)^{\alpha }Ge
n_{k\sigma \alpha}
] + { 1 \over 2} ce^2
+ \sum_{k,\alpha=1,2} (\Delta c_{k,\alpha,\uparrow}^{\dagger}
c_{-k,\alpha,\downarrow}^{\dagger} + h.c.) \nonumber \\
& &
  +bosonic
~terms
\ega
The first term corresponds to band-energy of the correlated
electrons in a two-fold orbitally degenerate $e_g$ band ; where
$U^\prime$ is the strength of the inter-orbital Coulomb
correlation treated within the mean field Hartree-Fock approximation, and the band
renormalization factor $\tilde q$ depends on the on-site intra-band coulomb interaction U. The
second term within the square bracket corresponds to the symmetry breaking electron -
lattice interaction that lifts the degeneracy of the bands by differently populating them and
thereby producing the lattice strain ($e$). The cost in lattice energy due to 
such strain is given by the second term whereas the
third term represents the usual BCS Cooper pairing between the quasiparticles of the correlated
systems in
which only intra-band pairing is considered for simplicity assuming the strength of pairing
interactions to be of
equal magnitudes \cite{7}. The bosonic terms in
the
Hamiltonian being C-numbers do not contribute to the dynamics of the
system.
In the normal undistorted phase, it is possible to show that the
strain is exactly proportional to the population difference
between the shifted sub-bands. The self-consistent 
equations for the strain ($e$) and the SC-order parameter($\Delta$) are
given below,

\bga
e  = -\sum_{k, \sigma, \alpha}(-1)^{\alpha}n_{k\sigma\alpha} 
 = -{G
\over c} \sum_{k, \sigma,\a}(-1)^{\alpha}{\ve_{\alpha}(k)
\over E_{\alpha }(k)} \tanh {{\beta E_{\alpha}(k)} \over 2}
\ega
\bga
\Delta  = -V \tilde q^2\sum_{k,\alpha} <c_{k\alpha \uparrow} c_{-k
\alpha
\downarrow} > 
=\sum_{k,\alpha}V\tilde q^2{\Delta \over {2E_{\alpha
}(k)}}\tanh{{\beta
E_{\alpha }(k)} \over 2}
\ega
where $\ve_{\alpha }(k)= \tilde q \epsilon_k - \mu + 2 U^\prime
<n_\beta>-
(-1)^{\alpha}Ge$ is the band energy of the correlated normal
state in the distorted phase and $E_{\alpha
}(k)=\sqrt{\ve_{\alpha }^2(k) +  \Delta^2}$ is the SC
quasiparticle energy in the band $\alpha $. 

In order to study the mutual influence of lattice distortion
and superconductivity as a function of band filling one has to
solve the above two coupled equations (3-4) self-consistently
together with the equation for dopant concentration which can be
obtained from the electron number conservation $ (2 - \delta )={1\over N}\displaystyle\sum_{k
\sigma \alpha }<n_{k\sigma\alpha}> $.
In order to solve these three equations the sum over $\vec k$ as usual is 
converted into an integral over energy variable $\epsilon$, with an approximate
peaked density of states $N_0(\epsilon)$ at the centre 
of the unperturbed bands as given below,
\begin{equation}
N_{0}(\epsilon)=N(0)\sqrt{1-\mid \frac{\epsilon}{B}\mid} ~ \ln
\mid {B^2 \over \epsilon^2}\mid
\end{equation}
\noindent where $N(0)$ is the unperturbed density of states of the free 
electron system and 2B (which is chosen as 1 eV) is the band width.

Figure 1. shows the effect of electron correlation on the
temperature
dependence of the strain, i.e, $e(T)$. In order to see this 
the curves for $e(T)$ corresponding to the value of the coupling constants
$G=1.2$
and band occupation $n=0.9$ are plotted for different values of u and $u^{\prime}$.
There is a suppression of strain around 40 K due to the appearance 
of superconductivity when u and $u^{\prime}$ are zero. As the intra-orbital intra-site Coulomb
correlation
is switched on with $u=0.2$, while keeping $u^{\prime}=0$, the
strain 
as well as $T_s$ has increased, and the SC transition temperature
is 
lowered to around 22 K. Furthermore, when the inter-orbital 
intra-site Coulomb interaction strength is increased from zero to
$u^{\prime
}=0.001$, there is a further suppression of $T_c$ and an
enhancement of $T_s$
as well as strain as can be seen from the dash - dotted curve. On
further 
increasing $u^{\prime}$ to 0.005 superconductivity almost vanishes
(dotted
curve) while $T_s$ and strain increases by a larger magnitude. Therefore, the present results
predict that the suppression of strain will be
lessened with increasing correlation in the superconducting state.

An unusual behavior in the temperature dependence of the SC order
parameter
$\Delta(T)$ and strain e(T) is observed when the value of the SC
transition temperature in the absence of strain ($T_c^{0}$) is
comparable 
to the structural transition temperature $T_s$, which is depicted
in Fig 2.
The figure depicts the temperature dependence of e and $\Delta$ for
the same
value of the coupling constant G=1.2eV as earlier but for a band
filling
$n=0.7$. In this case the strain appears at the temperature of $T_s
\approx 140 K$, increases with decreasing temperature but drops
precipitously
to zero around 92 K, as soon as superconductivity
appears. The SC order parameter rapidly increases within an
extremely narrow
range of temperature but it vanishes again with the recurrence of
strain 
which attains a large value when the temperature is reduced
slightly. Such
alterations of SC and strain persist until the temperature is
slightly less than 60 K below
which the superconducting state stabilizes and the strain vanishes.
This 
corresponds to a metastable situation that persists within a narrow
range of
temperature. While the SC state stabilizes at lower temperatures, the strain
exists in the
higher temperature range, with the two never co-existing in their
respective regions of stability. 

\par In figure 3, the occupation probability/per spin for different
orbitals are plotted as a function of temperature. The occupation probability of the two orbitals
are the same above the
structural
transition temperature $T_s$ but below $T_s$, the occupation
probabilities
being different creates a population difference between the orbitals
and 
hence a net strain builds up. With lowering of temperature, the
system 
undergoes SC transition and hence the electron density in each orbital is suppressed
below
$T_c$. Needless to say, this transition resembles to the
orthorhombic to 
tetragonal transition in high-$T_c$ systems. However,
as a result of reduction in the orbital population density below
$T_c$, 
the growth of the spontaneous strain gets arrested.



\par In figure 4, the detailed phase 
diagram comprising of $T_c$ $\&$ $T_s$ as a function of hole
concentration
is shown for different values of intra and inter orbital 
correlation strengths. The phase diagram (solid lines represent
$T_c$ curves)
demonstrates that the $T_c$ is largely suppressed in presence of
structural
distortion and the highest $T_c$ is obtainable for an optimum
doping 
($\delta_c$) where $T_s$ vanishes. It is to be noted that the suppression
in $T_c$ is maximum at half-filling as the strain is the largest there. These
results are in close agreement with experimental observations \cite{8,9}.

\par The structural transition in this model results from the
competition
between lowering of the electronic energy due to lifting of the
orbital 
degeneracy (band Jahn-Teller effect) and the increase in elastic
energy
associated with the apperance of strain. The gain in energy in
this process
stabilizes the structural transition and this gain is largest when the
Fermi level (FL)
in the undistorted phase lies at the center of a peak in the DOS, because in that case a large
number of electrons in higher energy states are transferred to lower energy states. 
This is realized in the present model at half filling of the degenerate band resulting in a splitting
due to JT effect by an amount 2 Ge which leaves the lower band almost fully occupied whereas
the upper
band is nearly 
empty. Thereby, it creates maximum 
population difference between the two shifted sub-bands resulting in
the
largest strain. Consequently, the DOS at
the FL gets
reduced drastically so that the system is no longer favorable for superconductivity. However, as
one moves away from half-filling of the bands (by doping holes) the FL
moves away
from the peak in the DOS resulting in a lesser gain in electronic energy which will tend to
reduce
the distortion as well as the structural transition temperature. As a result the DOS at the FL will
increase thereby favoring superconductivity. Thus
the SC-$T_c$ increases with increasing hole concentration.

\par On the other hand, the appearance of SC reduces the occupation
probability of the
bands 
exponentially due to the presence of the gap 2$\Delta $ in the
energy spectrum resulting in the reduction of the 
strain ($e$) below $T_c$. This explains the arresting of the growth of the strain at
$T_c$ and 
its suppression below $T_c$. Such a
picture becomes self-evident from Fig. 3. 
Our results presented in figures have close resemblence to experimental data
of high temperature superconductors \cite{3,8,9,10}.

\par Finally, within the present formulation the electron correlation 
has the
following effects. The band narrowing due to renormalizatin factor $\tilde q$ enhances the
density of
states which favors the Jahn - Teller splitting, while the pairing amplitude 
gets suppressed, due to the reduction of the strength of the interaction by a factor $\tilde q^2$. 
Inclusion of inter
orbital coulomb
correlation makes the J-T splitting asymmetric with respect to the
centre 
of mass of the 
original band and hence enhances the lattice strain but suppresses
SC further.
Therefore, the present model study helps in understanding the 
interesting
phenomenon of the interplay of structural 
transition with superconductivity in correlated systems like the 
cuprates. However, it remains to be seen how the antiferromagnetic spin fluctuations
 leading 
to d-wave pairing superconductivity 
is influenced by structural distortion. This is the subject matter for a future study.  \\

\noindent{\bf Acknowledgement} One of us (HNG) would like to thank the Brazilian
funding agency FAPERJ for providing the financial support (project no.
E-26/150.925/96-BOLSA). \\

{\bf Figure Captions} \\

\noindent {\bf Fig. 1} The thermal variation of the lattice strain ($e$) for $n =0.9$ and $G = 1.2$ in presence of on-site coulomb correlation (intraband ($u$)
as well as interband ($u^\prime$ is denoted as $u1$ in figures)). \\

\noindent {\bf Fig. 2} The temperature variation of the superconducting gap
and the stain for $n= 0.7$, for $G=1.2$. A metastable thermal regime from
around $60$ K - $90$ K is worth noticing (the solid line indicates the 
$\Delta_{sc}(T)$ and the dotted one $e(T)$.

\noindent {\bf Fig. 3} The temperature variation of the occupation 
probabilities of the splitted bands for $n= 0.9$, $G =1.2$ in presence of
the onsite intraband ($u$) as well as ($u1 \equiv u^\prime$) coulomb 
correlations. \\

\noindent {\bf Fig. 4} The complete phase diagram in terms of the structural
and superconducting transition temperatures ($T_s$, $T_c$) as a function of
hole concentration ($\delta$) for $G =1.16$. 
The curves with connected lines correspond to $T_c$ curves).

\end{document}